\documentclass[11pt,a4paper]{article} % 10pt is ignored!
\usepackage{jheppub}
\usepackage{graphicx,subfigure}
\usepackage{amsmath,amssymb}
\usepackage{multirow}
\newcommand{\be}{\begin{equation}}
\newcommand{\ee}{\end{equation}}
\newcommand{\bea}{\begin{eqnarray}}
\newcommand{\eea}{\end{eqnarray}}
\newcommand{\ba}{\begin{array}}
\newcommand{\ea}{\end{array}} 
\def\o{\omega}

\title{Thermal Mass and  Plasmino for strongly interacting Fermions }
\author[a]{Yunseok Seo,}
\author[a]{Sang-Jin Sin}
\author[a,b]{and Yang Zhou}

\affiliation[a]{Department of Physics, Hanyang University,
Seoul 133-791, Korea}
\affiliation[b]{ Center for Quantum
Spacetime, Sogang University, Seoul 121-742, Korea}

\abstract{  
We investigate  fermion self energy problem in the strongly coupled dense medium 
in holographic approach. 
By working out bottom up models as well as top down ones 
we showed that vanishing  thermal mass and non-existence of temperature generated 
plasmino mode is the universal feature of the strongly interacting fermion system. 
We identified that  the dual of the 
bulk Rashiba effect, which was recently found by the Herzog et.al, is the presence of the 
plasmino mode   generated by the density. 
}

\begin{document}
\maketitle
\section{Introduction}
One of the most tantalizing problem is to calculate the physical quantities of the strongly 
interacting system.  While standard field theory technique does not work,
 the gauge/gravity duality has some chance for such problem since it  maps a strongly interacting quantum system to the weakly coupled classical system. Actually it has been  applied to   variety of problems with many successes.
 In particular, strongly interacting fermion system is of particular interest due to its important role 
 in condensed matter as well as nuclear physics.  
In conventional field theory  approach to the fermion with {\it strong coupling}, there are two serious problems: 
one is the truncation of the infinite diagrams to the ladder type in solving the Schwinger-Dyson equation 
and the other is the presence of the chemical potential, which gives  the fermion sign problem in numerical investigation.
 
  Applying the  gauge/gravity duality  to   the  fermion   in strongly coupled system,    
 it was pointed out~\cite{Lee:2008xf} that 
the very existence of the fermi surface itself is a serious issue . 
By studying the   fermion spectral function~\cite{Liu:2009dm,Faulkner:2009wj,Cubrovic:2009ye}, it was  shown that the fermi surface quickly disappears as one increases  the temperature and non-fermi liquid nature is overwhelming in  strong interacting regime.  
 In~\cite{Herzog} the dispersion relation was studied in the presence of the chemical potential and it was pointed out that  there is an interesting spin orbit coupling  due to the bulk electric field dual to the boundary global charge. 
 It raises an interesting issue since the spin-orbit coupling, $\sim k\times \sigma\cdot E$   which exists in the dual gravity side does not have any obvious dual in the field theory side:  the bulk electric field is a dual of `global' charge 
and there is no such electric  field in the boundary side. Identifying the dual of such bulk Spin-orbit interacting is one of the goal of this paper. 
   
The study of fermion self-energy has a long history due to its fundamental importance in studying electronic as well as nuclear matter system. 
{\it For the weakly interacting} field theory like QED or QCD,  the fermion self energy in the hot medium can be characterized 
 by the thermal mass generation of order $gT$ and presence of the plasmino mode  \cite{Braaten:1990it}, 
  a collective mode whose dispersion curve has a minimum at finite momentum. 
 These are well established and recorded in standard text books\cite{lebellac,kapusta}.  
However,   recent investigation suggested that perhaps such results do not continue to hold in strong coupling regime. In  \cite{Harada:2008vk}, it was reported that the thermal mass does not increase with increasing coupling using the Schiwinger-Dyson equation. 
Furthermore, in  \cite{Nakkagawa:2011ci}, in was claimed that the fermion thermal mass actually vanishes 
in the strong coupling limit. 
 Recently we studied this issue using the holographic method  and concluded that indeed there is no thermal mass generation for strongly coupled massless fermion system and there is no plasmino mode for de-confined system either for massless fermion.  We also showed that, in confined phase,   density effect can generate 
  plasmino mode if density is  large enough. The results were obtained in a very specific model of hQCD, namely  
 a  top down approach using D4 background with probe D8 ${\bar D}8$ system.  So it is natural to ask whether 
 such result is a universal feature of strongly interacting fermion system 
 or a feature of specific  model.
   
In this paper,  we  study  the problem  further and  conclude 
 that absence of thermal mass of massless fermion is universal feature 
by  showing that  bottom up models   and top down  ones produce qualitatively identical results. 
We also identify the universal  condition for the existence of the plasmino mode to be the presence of the large enough  chemical  potential: 
\begin{itemize}
\item   for confining geometry, 
plasmino exists for large enough chemical potential regardless of the fermion mass
\item for deconfining geometry, plasmino exists only for massive fermion with large enough density
\item  plasmino is absent at zero density, regardless of fermion mass, temperature, and phases (confinement/deconfinement)
\end{itemize} 
Finally we propose that  such density generated plasmino mode in strong coupling is the dual of the 
bulk spin-orbit-coupling or Rashiba effect. We stress that the temperature generated plasmino modes, which exist 
in the weak coupling, do not exist in the strong coupling result. 

The rest of the paper goes as follows. 
In section 2, we give a brief review of the known results on thermal mass and plasmino. 
In section 3, we  study the holographic fermions in top down model using the D4-D8-${\bar D}$8. 
In section 4, we use the bottom up model to study the same goal and got qualitatively the same results. 
In section 5, we summarize and conclude.

\section{ Thermal mass and Plasmino in   field theory}
\vskip0.2cm
We begin by giving a brief summary of hard thermal loop (HTL) discussion of plasmino.  
%\vskip0.1cm
In relativistic thermal or dense fermionic systems, there are two types of fermionic excitations. One is the ordinary fermion with dressed dispersion relation and the other is the plasmino. Generally two branches appear both in QED plasma and QCD plasma, namely both for electron propagator and quark propagator. In a weakly interacting field theory, plasmino is a collective excitation due to finite temperature or finite density effect.
 The fermion propagator is written as
\be
G(p)= \frac1{\gamma\cdot p-m+\Sigma(p)},\ee
 where the self-energy $\Sigma=\gamma_\mu \Sigma^\mu$  
 can be evaluated by one-loop calculation
 %to give \begin{equation}
 %\Sigma^0 = -{m_f^2\over 8 p}\log \left({\omega+p\over \omega-p}\right)^2\ ,~\Sigma^i = {m_f^2 \over 2 p^2}p_i\left[1-{\omega\over 4p}\log \left
 %({\omega+p\over \omega-p}\right)^2\right] \end{equation}
in the leading  order $T^2$ and $\mu^2$~\cite{Sigma}.
The gauge invariant result is available   for the hard thermal loop approximation in which  mass $m$ can be ignored  compared with $T$ or $\mu$.  In this limit the  fermion propagator can be decomposed as 
\bea\label{decompose}
G&=&{1\over 2}(\gamma_0-\gamma_ip^i)/\Delta_+ 
+{1\over 2}(\gamma_0+\gamma_ip^i)/\Delta_- \ ,\\
\Delta_\pm&=& \omega\mp p-{m_f^2\over 2p}\left[\left(1\mp{\omega\over p}\right)\log\left({\omega+p\over \omega-p}\right)\pm 2\right] .
\eea 
Here $p=|p_i|$ and $m_f$  is the effective mass generated by the medium effect~\cite{Sigma}
\be m_f^2={1\over 8}g^2C_F\left(T^2+{\mu^2\over\pi^2}\right)\ ,\ee where $C_F=1$ for electron and $C_F=4/3$ for quark. 
Notice that both thermal and density effect generate the effective mass. 
Solving for the pole of the propagator
we will get two branches of dispersion curves $\omega=\omega_\pm (p)$ and $\omega_-$ is the one that describes the plasmino. Asymptotic forms of $\omega_{\pm}$ are given 
by 
\bea
p<<m_f &:& \quad \omega_\pm(p) \simeq m_f \pm \frac13 p \label{asym} +
\frac{1}{3 m_f} p^2 + \cdots\ , \\ 
p>>m_f &:& \quad \omega_\pm (p) \simeq p\ .
\eea
\begin{figure} %  figure placement: here, top, bottom, or page
   \centering
  \includegraphics[angle=0,width=0.5\textwidth]{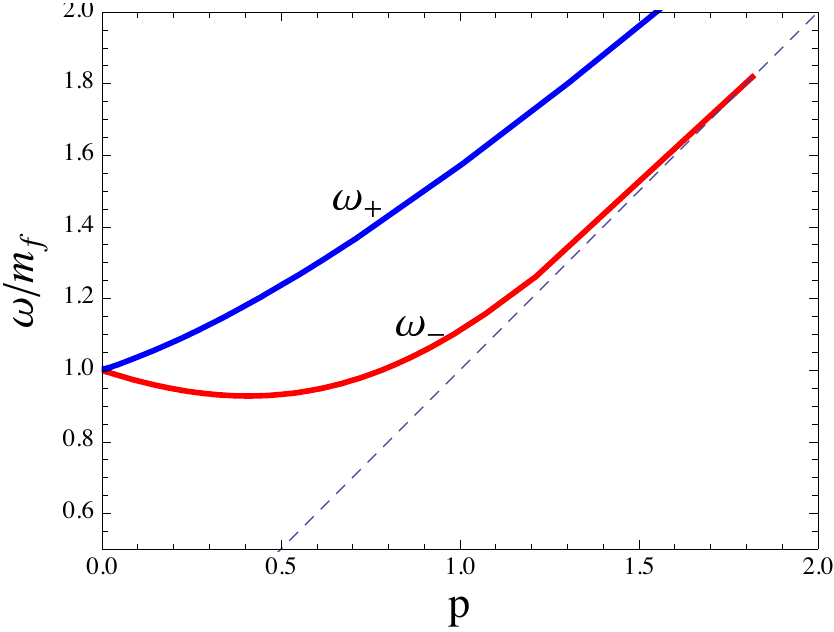}
   \caption{Two branches of dispersion relation for  fermion propagator. The dashed line is light cone.}
   \label{plasminoDispersion}
\end{figure}
Two branches are plotted in Figure \ref{plasminoDispersion}.  
%$\omega_+$ and $\omega_-$ are analog of transverse and longitudinal bosonic collective modes respectively. 
%Since (\ref{decompose}) is chiral invariant and pole $\omega_-$ and $\omega_+$ have opposite helicity, the $\omega_-$ excitation have negative %helicity over chirality ratio. 
The presence of the plasmino branch may be experimentally  important since it can enhance production rate of  the light di-lepton~\cite{Peshier:1999dt}. With such motivation, 
 plasmino has been investigated extensively~\cite{Braaten:1990it,Sigma,Weldon:1989ys,Pisarski:1989wb, Braaten:1991dd,Braaten,Baym:1992eu,Blaizot:1993bb,Pisarski:1993rf,Blaizot:1996pu,Schaefer:1998wd,
Peshier:1998dy, Peshier:1999dt,Mustafa:2002pb}. 
 For a review, see references~\cite{lebellac,kapusta}. 

In high temperature limit, fermion mass can be ignored therefore two branches   $\omega_\pm$ can be characterized in terms of chirality and helicity. The ratio of chirality and helicity are $\pm$ for $\omega_{\pm}$ respectively.
However,  in our approach, we do not neglect the fermion mass entirely therefore we define the plasmino character by the condition 
\be
\frac{d\omega_-}{dp} \Big |_{p \to 0} \cdot  \frac{d\omega_-}{dp}  \Big |_{p\to \infty} <0\ .
\ee
Notice that  plasmino in Hard thermal loop (HTL) approximation exists in high temperature whatever is the density.
However, in the strongly coupled region, we will show that plasmino disappears at zero density both in confined phase and deconfined phase and thermal mass vanishes there. Plasmino can survive only for a certain window of  chemical potential in confined medium. For deconfined phase, plasmino exists only for massive fermion with enough density. 
%Especially, at zero chemical potential, there is no plasmino. 
We will find that   the density and temperature independent  value $1/3$ in  (\ref{asym}) is  an artifact of HTL approximation. 

\section{Thermal mass and plasmino in Top Down hQCD}

\subsection{The Model}
We set up a calculational  scheme for fermion self-energy using  Sakai-Sugimoto (SS) model~\cite{Sakai:2004cn} 
where one uses black $D4$ geometry  and introduce a probe $D8/\bar{D}8$ for flavor dynamics.
 To introduce  density, we need  $U(1)$ gauge field on probe brane. 
 The source of the $U(1)$ gauge field are end points of strings which are emanating from   horizon  in deconfined phase and from  baryon vertex in confined phase.  
% This system is considered as a dual description of strong coupled finite dense medium at boundary theory. 

The geometry of confined phase is given by 
\bea
 ds^2 &=& \left({U\over
R}\right)^{3/2}\left(\eta_{\mu\nu} dx^\mu dx^\nu + f(U)dx_4^2\right) 
+ \left({R\over U}\right)^{3/2}\left({dU^2\over f(U)} + U^2
d\Omega_4^2\right)\ , 
\eea 
where both time and Kaluza-Klein direction are periodic:  $x^0\sim x^0+\delta x^0$, 
$x_4\sim x_4+\delta x_4$ and 
$$ f(U) = 1-U_{KK}^3/U^3\ ,~~ U_{KK} =
\left({4\pi\over 3}\right)^2 {R^3\over \delta x_4^2} $$
with  $R^3=\pi g_s N_c(\alpha')^{3/2}$. 
The solution also contains a nontrivial dilaton and a four form field;
$$ e^\phi =g_s \left({U/R}\right)^{3/4}\ , F_4= {\epsilon_4( 2\pi)^3(\alpha')^{3/2}N_c/ \Omega_4}\ .$$
Here following the original  Sakai-Sugimoto model, we consider the flavor eight brane embedded in $D4$ background 
with the
induced metric 
\be g_8^2 = \left({U\over R}\right)^{3/2}\left(\eta_{\mu\nu} dx^\mu dx^\nu +f(U)x_4'(U)dU^2
\right) + \left({R\over
U}\right)^{3/2}\left({dU^2\over f(U)} + U^2 d\Omega_4^2\right)\ .
\ee 
The action for the $U(1)$ gauge potential $A_0$ on eight brane  can be
obtained from the DBI action
 \bea 
S_{D8} &=& -T_8 \int d^9 x e^{-\phi} \sqrt{-\det(g_{MN} + 2\pi \alpha' F_{MN})}\ . \\
&=& N  \int dr~ r^4 \sqrt{f(r) (x'_4)^2 + r^{-3}\left({1/ f(r)} - a_0^{\prime 2}\right)}\ ,\\
&=& N  \int dr~ r^4 \sqrt{r^{-3}\left(1/ f(r) - a_0^{\prime 2}\right)}\ ,
\eea 
where we consider the trivial embedding at antipodal points $x_4'(r)=0$. 
The overall normalization $N$ is defined by 
$ N:={T_8\Omega_4 v_3\delta x^0 R^5/g_s}\ ,$ and $\Omega_4$ is the unit four sphere volume and $v_3$ is the three Euclidean space volume. 
For later convenience, we use dimensionless quantities $$ r = U/R\ ,~r_0=U_{KK}/R\ ,~a_0 = 2\pi \alpha' A_0/R\ . $$ The first integral   for $a_0(r)$ is 
\bea\label{a0eqc}
\frac{ra_0^\prime(r)}{\ \sqrt{r^{-3}\left({1/ f} - a_0'(r)^2\right)}} &=& D\ ,\eea 
 where $D$ is an integral constant representing the charge density.
  We can solve the gauge field profile
$a_0(r)$ by imposing boundary condition.

The deconfined geometry is given by
 double Wick rotating $x^4$ and time from the confined one
 %So we do not write it down explicitly. 
\be   ds^2 = \left({U\over R}\right)^{3/2}\left(-f(U) dt^2+ dx_i dx^i + dx_4^2\right) + \left({R\over U}\right)^{3/2}\left({dU^2\over f(U)} + U^2 d\Omega_4^2\right)\ . \ee 
%and also 
The gauge field profile on eight brane is solved as before by replacing $D\to D'$ and $f\to 1$  in (\ref{a0eqc}):
\be\label{a0eqd}
{ra_0^\prime(r)\over \sqrt{r^{-3}\left(1- (a_0'(r))^2\right)}} = D' \ .
\ee  The difference of two first integrals (confined and deconfined phases) is due to the presence of $f(r)$ factor. 
Chemical potential can be defined as the value of $a_0(r)$ at the infinity when we specify the boundary condition at IR.
 In deconfined phase we set $a_0(r_H)=0$ at the horizon while, for confinined case, we choose to set $a_0(r_0)=m/q$ in order to include the baryon mass so that the chemical potential $\mu$ is given by  
 \be\label{totalmu}
 \mu={m\over q}+\int_{r_0}^\infty a^\prime_0(r)\  dr.
 \ee\label{conchemical}
% We will determine $m_*$ in a self-consistent way later. 
This baryon mass is naturally related to the five dimensional Lagrangian fermion mass $m_5$ as we will see later.  
%$m_5$ will be related 
%to the actual 4 dimensional vacuum fermion mass $m$ later. 
% In confined phase $a_0(r_0)$ is zero or the mass of the baryon vertex depending on the absence or presence of the latter[S~~].
\subsection{Fermionic Green function on D brane world volume}
Now we study the fermionic excitation in the holographic dual background. Consider the probe fermion field in the world volume of D8 brane.
We  ignore the $S^4$ dependence and work in the effective 5 dimensional world volume  following the original paper of SS model.
%with the {\it induced 
%metric}  written as 
%\be
%ds_5^2 = R^2\left(r^{3/2}\left(\eta_{\mu\nu} dx^\mu dx^\nu \right) + r^{-3/2}{dr^2\over f(r)}\right)\ . \ee 
%The metric
%component is given by \be g_{tt} = R^2 r^{3/2}\ , ~g_{ii} = R^2 r^{3/2}\ ,~g_{rr} = R^2\left(r^{3/2}fx_4' +
%r^{-3/2}f^{-1}\right)\ . \ee 
%The tetrad are given by 
% \be e^{\bar t}_t = R r^{3/4}\ ,~
%e^{\bar i}_i = R r^{3/4}\ , ~e^{\bar r}_r = R\sqrt{
%r^{-3/2}f^{-1}}\ , \ee 
%where bar signifies indices in the tangent
%space. 
We use the minimal action \be S =
\int d^5x \sqrt{-g}\  i \left(\bar\psi\Gamma^M D_M\psi -
m_5 \bar\psi\psi\right)\ , \ee where the covariant derivative is $D_M = \partial_M + {1\over 4} \omega_{abM} \Gamma^{ab} - i q A_M$. $M$ denotes the bulk spacetime index while $a,b$ denote the tangent space index. By a factorizing  
$$
\psi=(-gg^{rr})^{-1/4}e^{-i \omega t + i k_i x^i}\Psi\ ,
$$ the Dirac equation can be give by
\be\label{dirac1}
\sqrt{g_{ii}/ g_{rr}}(\Gamma^{\underline{r}}\partial _r - m_5\sqrt{g_{rr}})\Psi + i K_\mu \Gamma^{\underline{\mu}}\Psi = 0\ ,
\ee where $K_\mu = (-v(r), k_i)$ and     $v(r)= \sqrt{-g_{ii}/g_{tt}} (\omega + q a_0)$.
Following the procedure in~\cite{Faulkner:2009wj},
we rewrite the Dirac equation (\ref{dirac1}) in terms of  two component spinors 
$\Psi=(\Phi_1,\Phi_2)$  in a decoupled form
\be\label{EOMM}
(\partial_r + m_5 \sqrt{g_{rr}}\sigma^3)\Phi_\alpha = \sqrt{g_{rr}/ g_{ii}}(i\sigma^2 v(r) +(-1)^\alpha k\sigma^1)\Phi_\alpha\ ,
\ee where $\sigma^i$ are Pauli matrices and $\alpha=1,2$.
Decomposing $$\Phi_1=(y_1, z_1)^{T}\ ,~\Phi_2=(y_2, z_2)^{T}\ ,$$ we get equations of motion 
for the component fields. For $(y_2,z_2)$ we get
\begin{eqnarray}
(\partial_r + m_5 \sqrt{g_{rr}})y_2(r) &=& \sqrt{g_{rr}/ g_{ii}}(+ v(r)+k)z_2(r)\ ~\label{wave1}\\
(\partial_r - m_5 \sqrt{g_{rr}})z_2(r) &=& \sqrt{g_{rr}/g_{ii}}(-v(r)+k)y_2(r)\ .~\label{wave2}
\end{eqnarray}
By replacing $k$ by $-k$, we obtain the equations of motion for $y_1$ and $ z_1$. Retarded Green function can be expressed in terms of variables
% We will see that this transformation does not change chirality.
 %We first consider the massless equations of motion for $\Phi_2$. Near the UV boundary we found
% \be
% y_2(r)\rightarrow \hat y\ ,\quad\quad  z_2(r)\rightarrow \hat z\ ,
% \ee  which define a ``source" $\hat y$ and  the corresponding ``expectation value" $\hat z$. This provides us to define the retarded Green function \footnote{In massive case, we need a non-trivial normalization for the Green function.}. Generally we can define a $r$ dependent Green function
  $$ G_1(r) : = {y_1(r)\over z_1(r)}\ ,~
 G_2 (r) := {y_2(r)\over z_2(r)}$$ as 
 \be 
 G^R_{1,2} = \lim_{\epsilon\rightarrow 0} e^{8 m_5Rr^{1/4}} G_{1,2}(r)|_{r=1/\epsilon}\ ,
 \ee where $G_1$ and $G_2$ satisfy the following equations
  \begin{eqnarray}
\sqrt{g_{ii}\over g_{rr}}\partial_r G_\alpha + 2m_5\sqrt{g_{ii}}G_\alpha =(-1)^\alpha k +
v(r)+\left((-1)^{\alpha -1} k+ v(r)\right)G_{\alpha}^2\ . ~~\label{floweq2}
\end{eqnarray}
Now we want to solve  (\ref{floweq2}) by imposing proper boundary conditions. 
In the confined phase, $v(r)= \omega + q a_0(r)$
%, and $g_{rr} = R^2 r^{-3/2}f(r)^{-1}$.
% The tetrad are as follows \be e^{\bar t}_t = R r^{3/4}\ ,~
%e^{\bar i}_i = R r^{3/4}\ , ~e^{\bar r}_r = R\sqrt{r^{3/2}fx_4' +
%r^{-3/2}f^{-1}}\ , \ee where bar signifies indices in the tangent
%space. 
%  For the trivial embedding
%  function $x_4'(r)=0\ ,$ we obtain
%  \be
%g_{rr}= R^2\ r^{-3/2}f^{-1},\   
%and
%$v(r)= \omega + q a_0(r)$, 
%  \ee  
  % \be
%{D^2\left(f(x_4')^2 + r^{-3}f^{-1}\right)\over (r^2+D^2r^{-3})} 
%={D^2\left(r^{-3}f^{-1}\right)\over (r^2+D^2r^{-3})}=(a_0')^2
%  \ee 
with 
  \be
a_0(r)= \mu + \int_{\infty}^{r} d\hat r \sqrt{D^{2} \over (\hat r^5 +D^{2})f
}\ .
  \ee
  Notice that $g_{rr}$ diverges at $r_0$.  For the  regularity of (\ref{floweq2}), we impose following boundary condition 
   \be\label{bcondition1}
G_\alpha(r_0) =  {-m  R  +\sqrt{m^2 R^2 +k^2- {\hat \omega}^2}\over (-1)^\alpha k- \hat\omega}\ ,
%G_1(r_0) ={m-\sqrt{m^2+k^2-\omega^2}\over k+\omega}  
\ee 
where $\hat\omega=\omega+m$ and 4 dimensional vacuum mass of the fermion is defined as $m:=m_5r_0^{3/4}$. 
%where $\hat\omega=\omega+qa_0(r_0)$.
% In massless limit, IR boundary conditions become
 %\be\label{bcondition1m}
%G_2(r_0) = \sqrt{k+\omega\over k-\omega}\ , \quad\quad G_1(r_0) =-\sqrt{k-\omega\over k+\omega}\ .\ee
Imposing the boundary condition for retarded (advanced) Green function corresponds to $$\omega\rightarrow\omega + i \epsilon\ ,~(\omega\rightarrow\omega - i \epsilon).$$
In the deconfined phase $$v(r)= {\left(\omega + q a_0(r)\right)\over \sqrt{f}}\ ,~~
a_0(r)= \mu + \int_{\infty}^{r} d\hat r \sqrt{D'^{2} \over \hat r^5 +D'^{2}\ .
}$$
The IR boundary condition in this case is 
\be\label{bcondition2}
G_{1,2}(r_H) = i\ ,
\ee required by the  in-falling condition  at the black hole horizon.

%\newpage
\subsection{Confined Phase}
We discuss the results of dispersion relation by solving (\ref{floweq2}). The dispersion relations for excitations in confined phase are reported in this subsection and we will discuss those in deconfined phase in next subsection.
\subsubsection{Light cone structure}
We first study the relation between dispersion relation and light cone structure. For simplicity, we set $m_5=0$ but things are similar in the case of finite mass as we will discuss later. \par
\begin{figure} [ht!]
\begin{center}
\subfigure[]{\includegraphics[angle=0,width=0.35\textwidth]{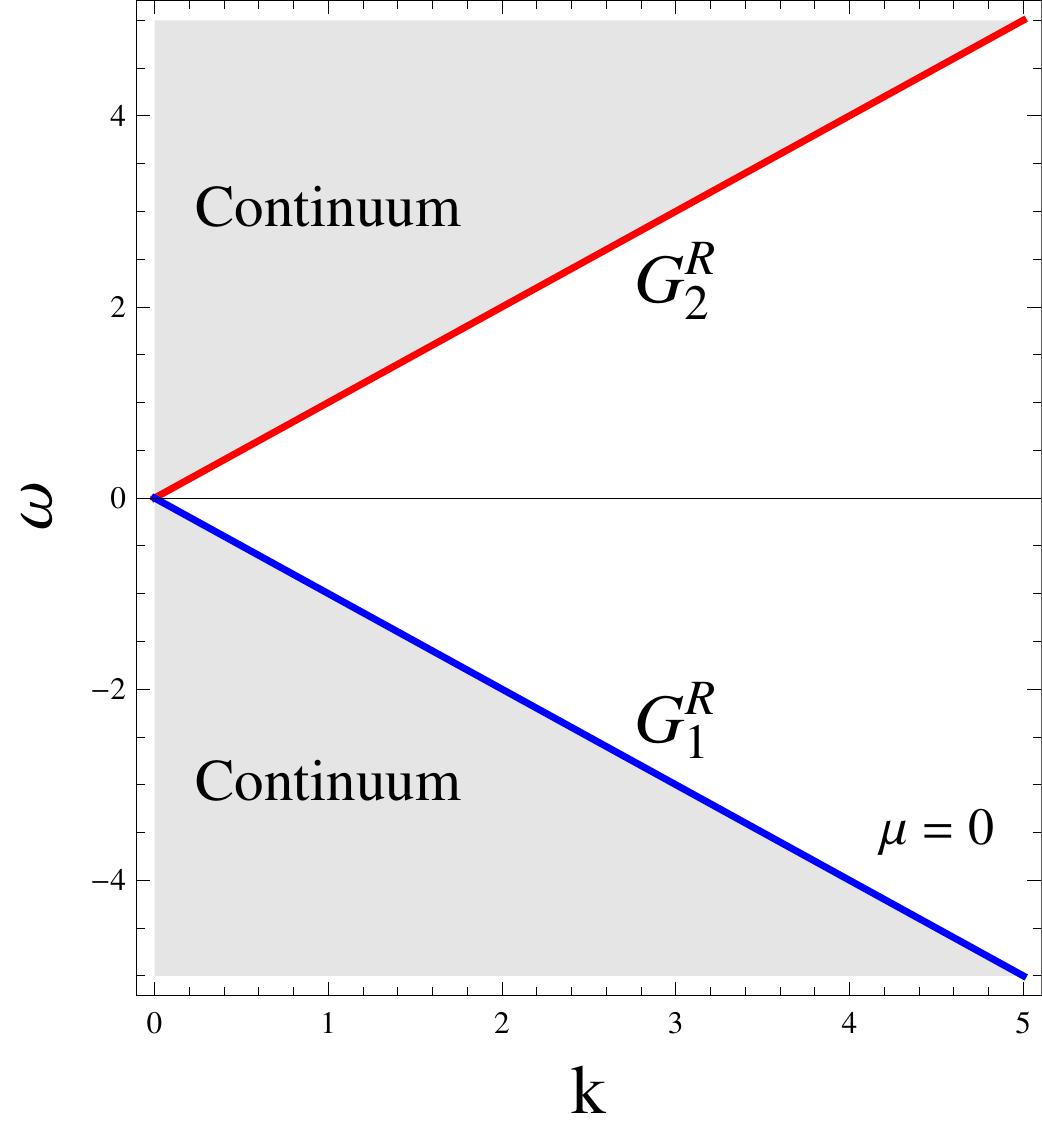}} 
\hspace{1cm}
\subfigure[]{\includegraphics[angle=0,width=0.38\textwidth]{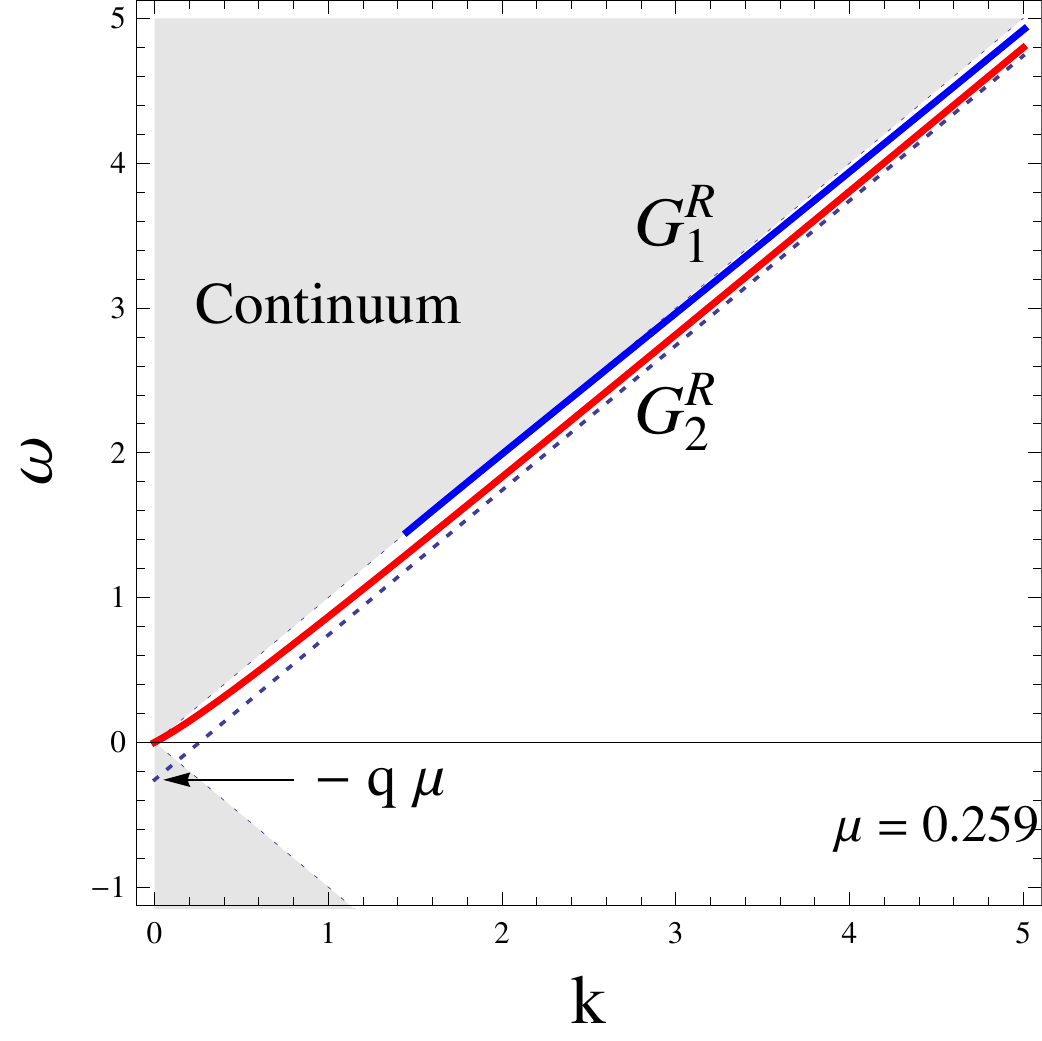}}  
\caption{(a)Dispersion curve without chemical potential. (b) Dispersion curve with $\mu =0.259$. Dashed line denotes UV light cone.
\label{fig:disp01}}
\end{center}
\end{figure}

%\cite{Allais:2012ye}{\it (We need to add this paper in proper place)} \par
When $m_5=0$, the IR boundary condition for retarded Green's function can be obtained from 
(\ref{bcondition1}): 
\be\label{bc01}
G_2 (r_0) = \sqrt{\frac{k+(\o +i \epsilon)}{k-(\o+i \epsilon)}},~~~~~ G_1 (r_0) = -\sqrt{\frac{k-(\o+i\epsilon)}{k+(\o+i \epsilon)}} = -\frac{1}{G_2 (r_0)}.
\ee
The IR boundary values contain two features. 
First, %the  spectral density is defined as an imaginary part of IR Green's function:
%\be
%{\rm Im}(G_{1,2}) =\sqrt{\frac{  \o^2 +\epsilon^2-k^2 + \sqrt{( \o^2+\epsilon^2-k^2)^2 + 4 k^2 \epsilon^2}}{2\left( (\o \pm w)^2 +
%\epsilon^2\right)}},
%\ee
%in denominator, $+$ sign is for $G_1$ and minus sign is for $G_2$.  It has sharp peak at $k =\pm w$ and it's hight behaves like
%\be
%{\rm Im}(G_{1,2})\Big|_{k = \pm \o} \sim \sqrt{\frac{k}{\epsilon}}.
%\ee
%In the limit $\epsilon \rightarrow 0$, it gives delta function peak. 
%This result is also consistent with numerical calculation in  Figure \ref{fig:disp01}(a).
%It means that 
the dispersion relation coming from IR boundary value is light-like and we call $k=\pm \o$ lines as IR light cones. Notice that when $m=m_5r_0^{3/4}=0$, chemical potential vanishes at the IR boundary following (\ref{conchemical}). This is always true even we consider finite chemical potential at UV. Therefore IR light cones do not change as we change chemical potential.
Second,  IR boundary value of $G_2$($G_1$) in (\ref{bc01})  becomes pure imaginary when $ \o>k $ ($\o<-k$) for positive $k$.
%In this case, spectral density at UV boundary has finite height. 
We refer this ($\omega, k$) region to continuum region since there is no definite peak~\cite{Allais:2012ye}. See for example, figure~\ref{fig:disp01}. 

Actually (\ref{floweq2}) implies how Green's function flows from  IR to UV. If we turn off gauge field everywhere,
the IR condition (\ref{bc01}) itself is a solution of (\ref{floweq2}) and Green's function does not run for zero chemical potential. Therefore UV dispersion relation we are interested in is light-like as the IR one. 

\begin{figure} [ht!]
\begin{center}
\includegraphics[angle=0,width=0.35\textwidth]{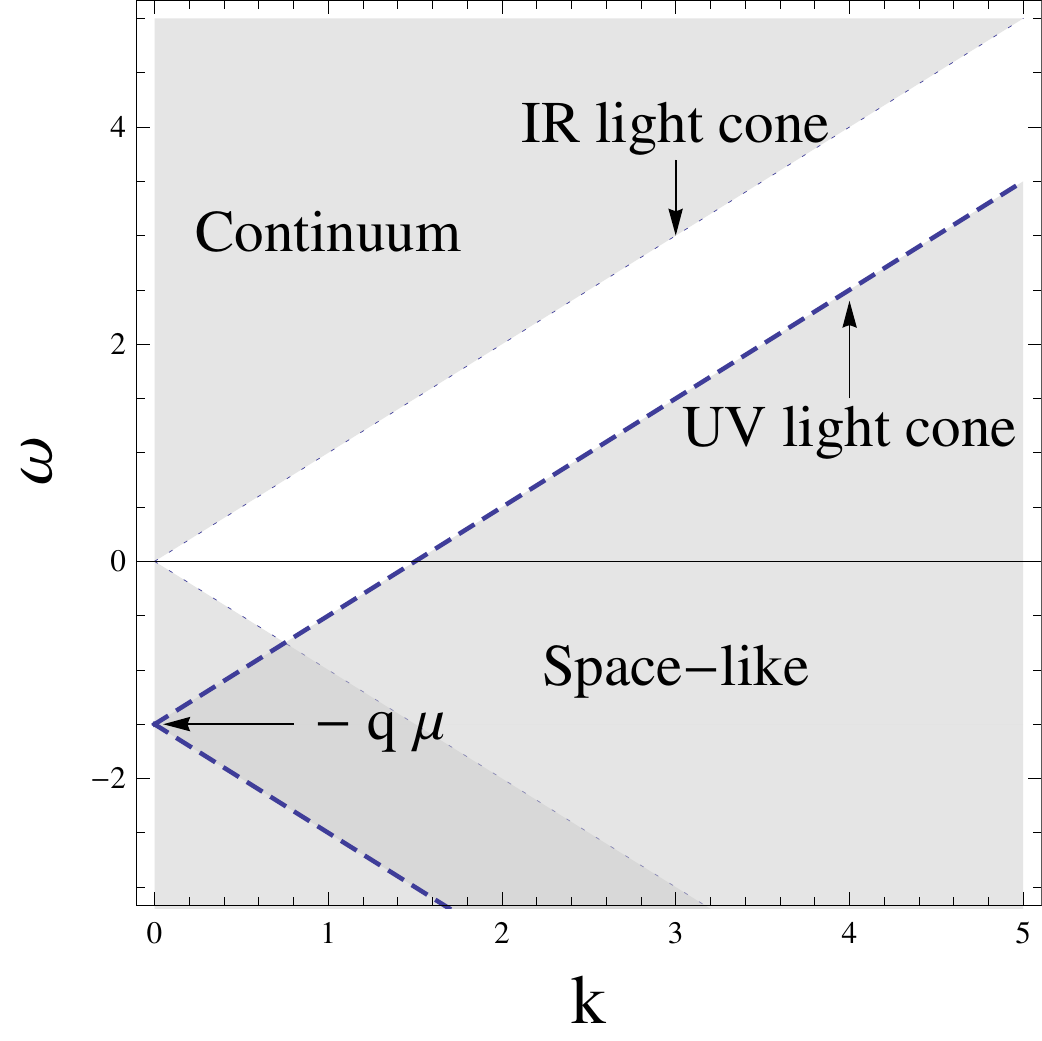}
\caption{UV and IR light cone with finite chemical potential. Dispersion curve should be located in unshaded region.
\label{fig:iruvlc}}
\end{center}
\end{figure}

However, if we turn on chemical potential, there is a drastic change in the dispersion curve (UV). The dispersion curve for $G_2^R$ deviates from IR light cone as we increase chemical potential, but the one for $G_1^R$ (located at $\omega=-k$ previously for zero chemical potential) immediately disappears and a new one appears as we can see from Figure \ref{fig:disp01} (b). This behavior can be explained by  causality condition. At zero chemical potential, two dispersion curves  are located on the light cones. If we turn on  chemical potential, the UV light cone is shifted down by $- q \mu$ so that  the previous negative branch $\omega=-k$ (with positive $k$) now lies in the causally forbidden region of the new light cone. See Figure \ref{fig:iruvlc}.
 The dispersion curves now can only appear inside the new light cone and outside of the original one (IR one), which is  the unshaded region of the Figure \ref{fig:iruvlc}.

\subsubsection{Dispersion relation for massive fermion}
\begin{figure} [ht!]
\begin{center}
\subfigure[]{\includegraphics[angle=0,width=0.35\textwidth]{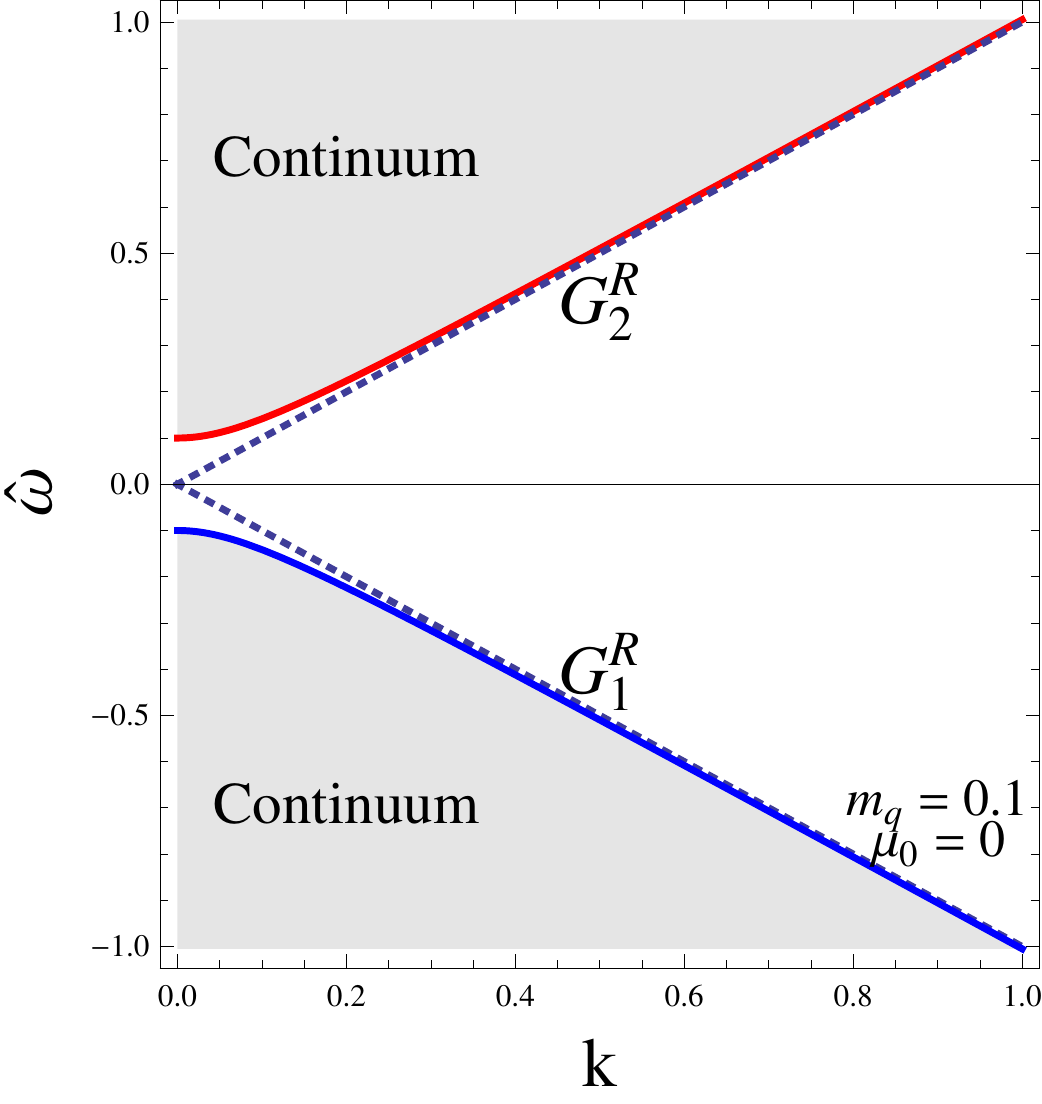}}
\hskip.7cm
\subfigure[]{\includegraphics[angle=0,width=0.4\textwidth]{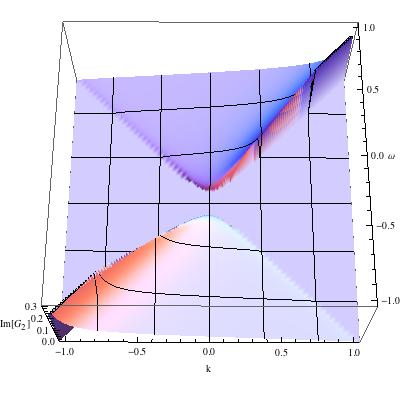}}
\caption{(a) Continuum region for $m_q = 0.1$ and blue (down) and red (up) solid lines represent the dispersion curves at zero chemical potential for $G^R_1$ and $G^R_2$ respectively. Dashed lines are light cones. (b) 3D plot of Imaginary part of $G_2^R$ at zero chemical potential. \label{fig:continuum}}
\end{center}
\end{figure}

The baryon vertex  is connected to the probe branes by the $N_c$ strings. 
From the probe brane point of view, the baryon is a collective state of the end points of the strings whose individual 
entities are the quarks. As  usual, the end points are also the source of the U(1) gauge fields. 
   The value of gauge field at IR boundary is identified as the mass of baryon vertex. 
   The collective motion of the end points is presented by a fermion field describing baryons living on the probe brane.   
   In this work we set mass of fermion equal to mass of baryon vertex.
   In this picture baryons are always massive and of order $N_c$. 
The equation of motion for the Green function is given by (\ref{floweq2})
with  boundary condition 
\be\label{bc2}
G_\alpha(r_0) =  {-m +\sqrt{m^2  +k^2- {\hat \omega}^2}\over (-1)^\alpha k- \hat\omega}\ .
%G_1(r_0) ={m-\sqrt{m^2+k^2-\omega^2}\over k+\omega}  
\ee 
Notice that we set $R=1$ to simplify the discussion. Due to the mass term in the square root, the continuum region is changed accordingly:
\be
\hat{\omega} > \sqrt{k^2 +m^2},~~~~\hat{\omega}< - \sqrt{k^2 + m^2}
\ee
which is shown in Figure \ref{fig:continuum} (a).

\begin{figure} [ht!]
\begin{center}
\includegraphics[angle=0,width=0.5\textwidth]{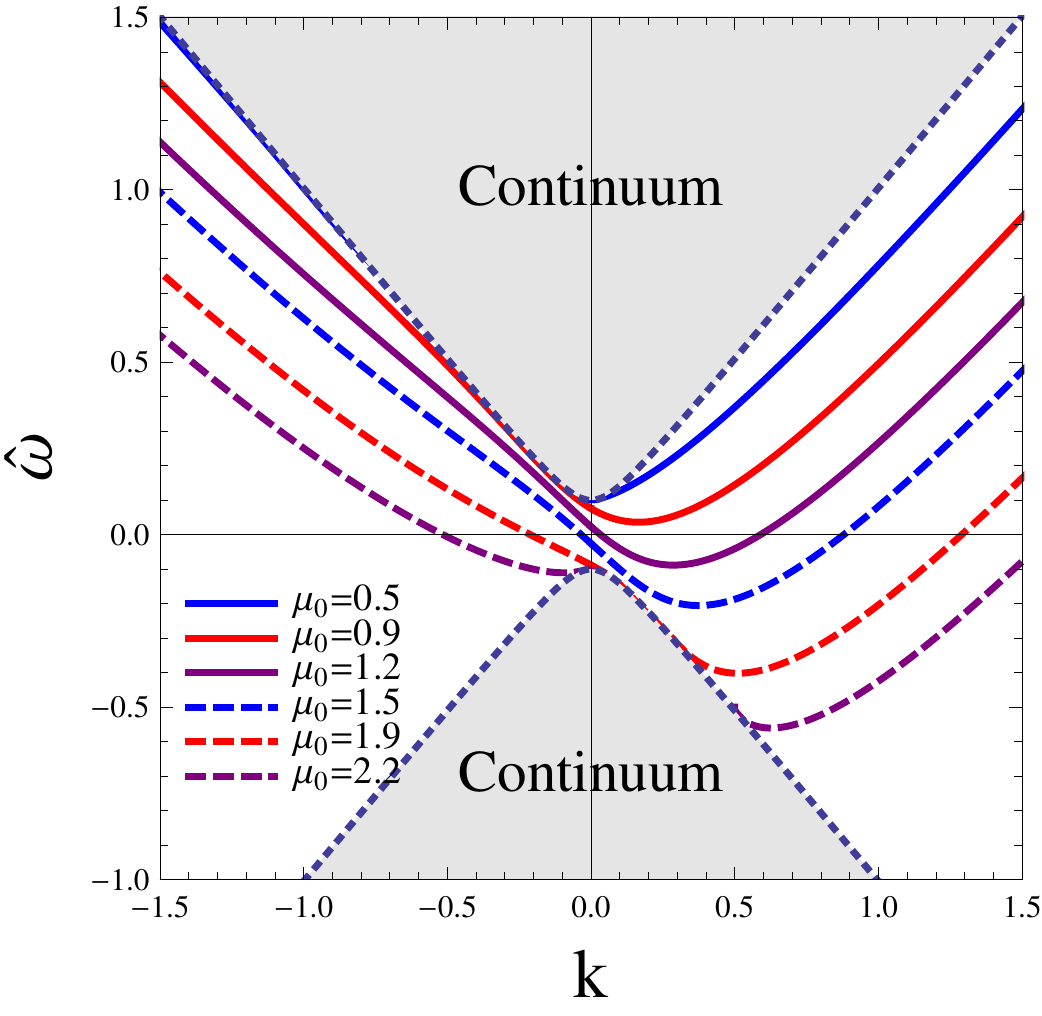}
\caption{ Dispersion curve of $G_2^R$ with several values of chemical potential in confining geometry. Here $\mu_0$ does not contain mass term. 
\label{fig:G2_mu}}
\end{center}
\end{figure}

%(\ref{floweq2}) can be solved numerically by imposing boundary condition at $r=r_0$ (\ref{bc2}). 
In the case of zero chemical potential, the poles are located at the edge of continuum region. See Figure \ref{fig:continuum}(b).  
As we increase chemical potential, dispersion curve for $G^R_2$ moves down.  The chemical potential dependence of dispersion curves of $G^R_2$ is drawn in Figure \ref{fig:G2_mu}, where
we can see several interesting behaviors.
 When chemical potential is larger than a certain value  $\mu_1 \sim 0.58$, slope at $k=0$ becomes negative, therefore plasmino appears. As we increase chemical potential further, two branches ($k<0, k>0$) are smoothly connected  at $\mu_0=0.9$, see the solid red line in Figure \ref{fig:G2_mu}. At $\mu_2 \sim 1.7$, the dispersion curve touches the edge of lower continuum region.
 If we increase chemical potential further, the dispersion curve splits again and moves down along the edge of lower continuum region. See the dashed purple line in Figure \ref{fig:G2_mu}.
 
\begin{figure} [ht!]
\begin{center}
\subfigure{\includegraphics[angle=0,width=0.3\textwidth]{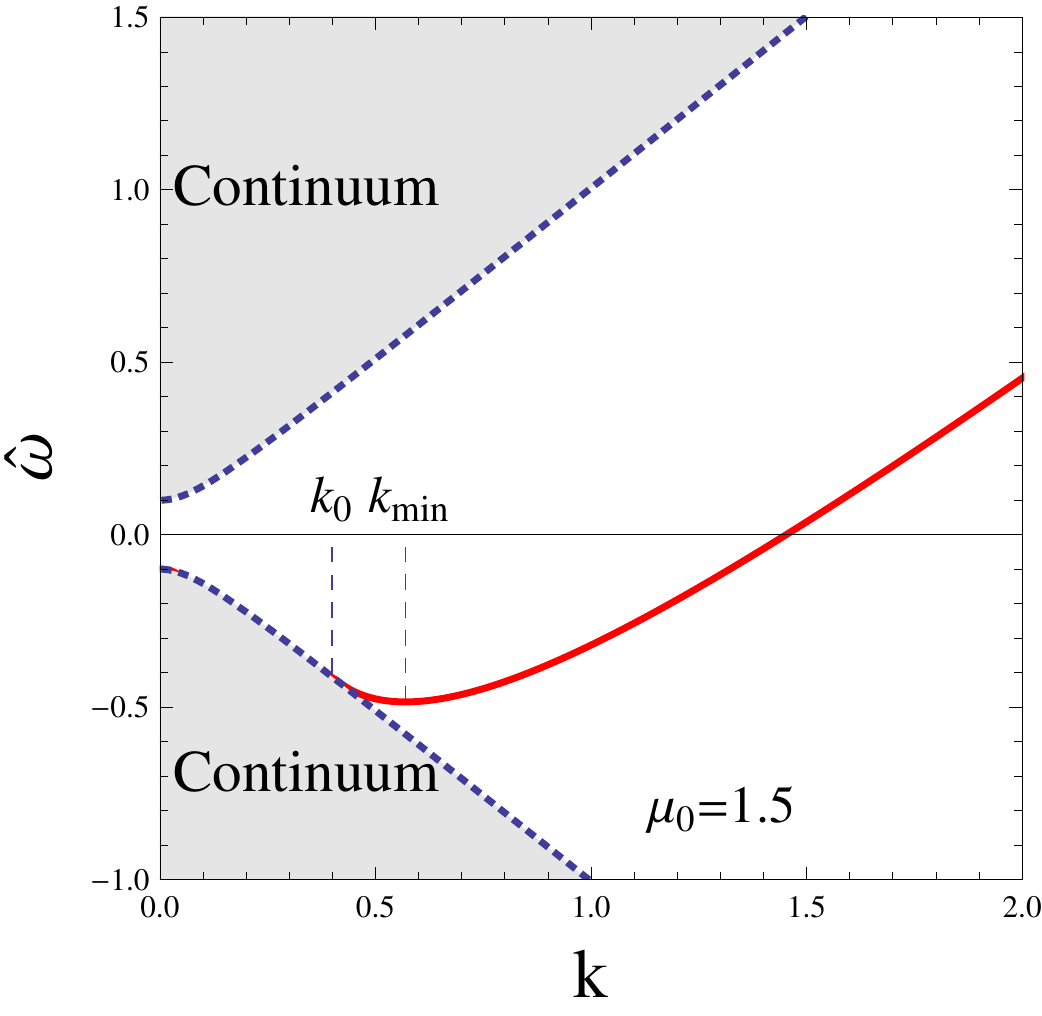}}
\hskip1cm
\subfigure{\includegraphics[angle=0,width=0.4\textwidth]{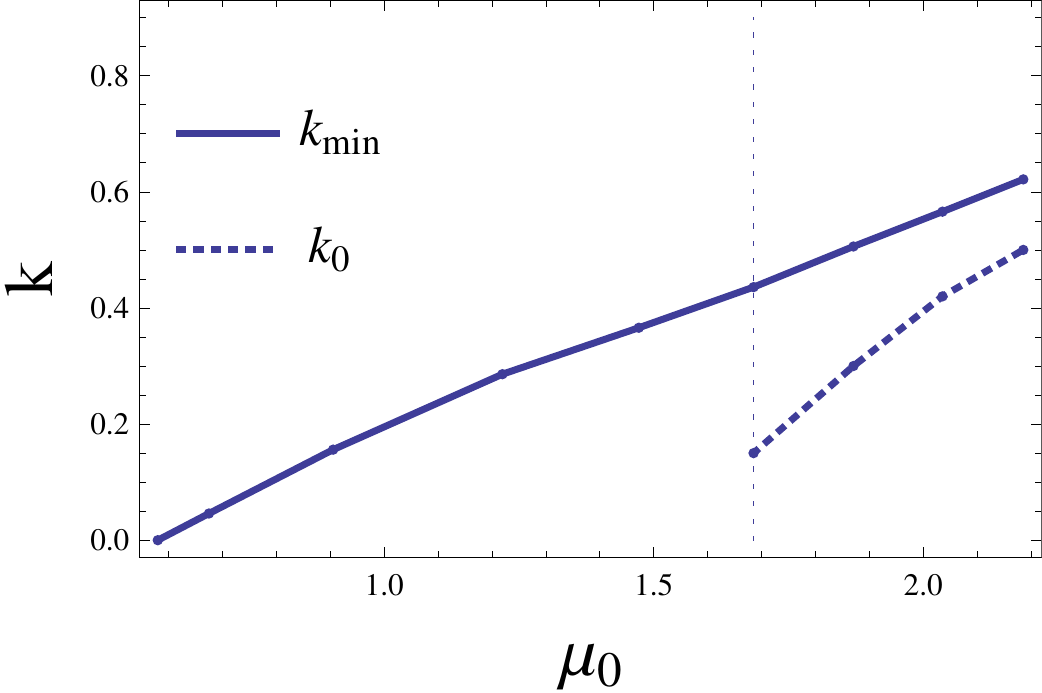}}
\caption{(a) Definition of  $k_{\rm min}$ and $k_0$ in dispersion relation of $G_2^R$ . (b) Chemical potential dependence of $k_{min}$ and $k_0$.
\label{fig:kokmin}}
\end{center}
\end{figure}

%From the dispersion relation, we can get several properties of the system.  If chemical potential is larger than $\mu_1$, slope of dispersion curve at $k=0$ becomes negative and  %plasmino behavior appear. The dispersion curve at large momentum approach $\omega = k -\mu$ which has positive slope. Therefore, 
There exists a certain momentum where the slope of dispersion curve $d \o /d k$ vanishes. We denote this momentum $k_{min}$. If chemical potential is larger than a certain value, say $\mu_2$, dispersion curve touches lower continuum region. We call the value of this momentum as $k_0$, see Figure \ref{fig:kokmin}(a). The density dependence of these values are drawn in Figure \ref{fig:kokmin}(b). 

\begin{figure} [ht!]
\begin{center}
\subfigure[]{\includegraphics[angle=0,width=0.3\textwidth]{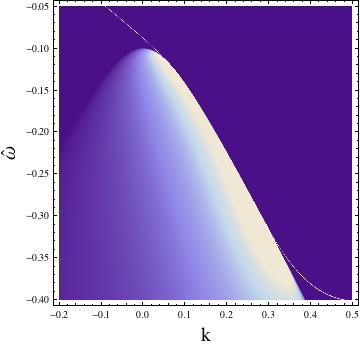}}
\hskip1cm
\subfigure[]{\includegraphics[angle=0,width=0.35\textwidth]{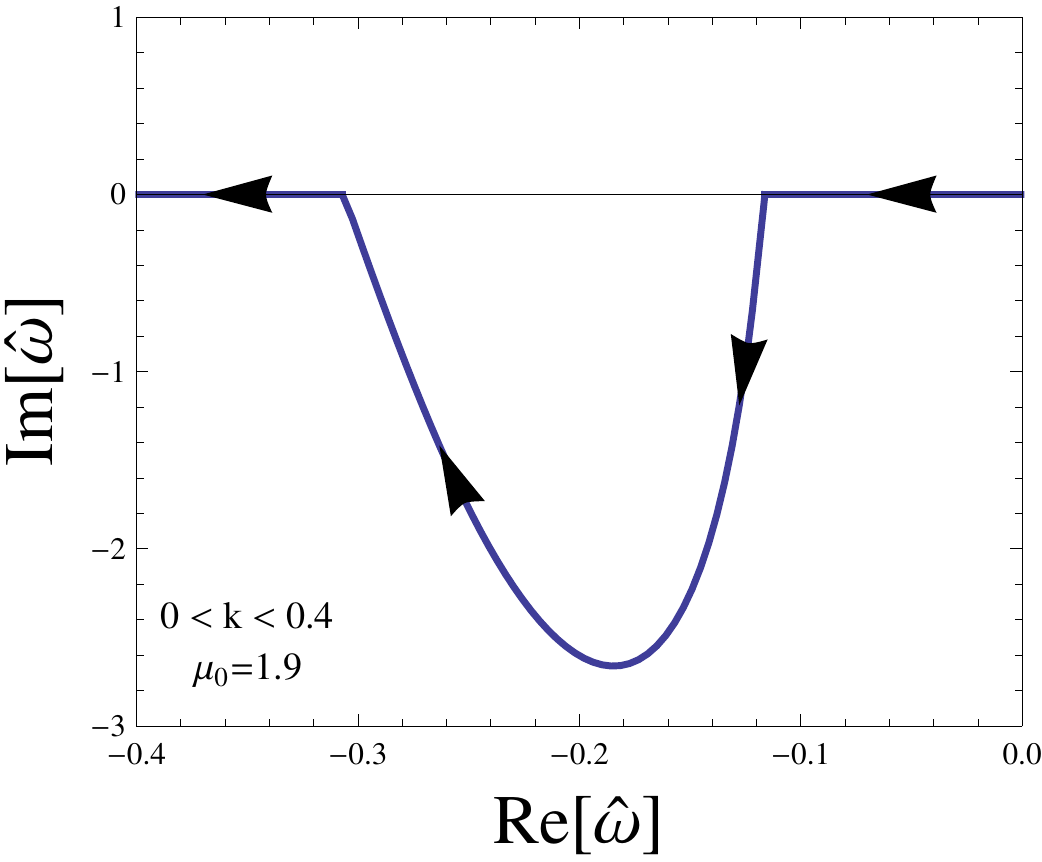}}
\caption{(a) Density plot of spectral function with $\mu_0 = 1.9$. (b) Position of $\omega$ in complex plane in the range of $0<k<0.4$ with $\mu_0 =1.9$. Arrows denote increasing momentum. \label{fig:G2dens}}
\end{center}
\end{figure}

In the region where chemical potential is larger than $\mu_2$, 
dispersion curve develops unstable part between $k_1<k<k_0$ where the pole in $\omega$ plane  develops imaginary part
(equivalently self energy develops imaginary part) 
and the residue $Z(k)$  of the pole develops singularity near $k_0$, so that spectral function develops big mountain there.  
This behavior of Green function suggests that for very high density, 
the low momentum part of the plasmino mode is unstable with large decay width.   
The low momentum instability of the quasi particle also  exists  in normal mode as well as in plasmino mode. 
 See figure \ref{fig:G2dens}. 
 
In Figure \ref{fig:G2dens}(a), we zoom up the lower continuum region where dispersion curve is touching. The behavior of $\omega$ in complex plane is drawn in Figure \ref{fig:G2dens}(b) in the range of $0<k<0.4$. In $0<k<k_1 =0.06$ region, the peak is infinitely sharp which means that $\omega$ does not have imaginary part, it corresponds to $-0.1<\omega<0$. Once dispersion curve touches the continuum region, the peak of spectral density becomes lower and broader, and the imaginary part of $\omega$ starts to develop in this region($0.06 < k < 0.29)$. If the momentum is larger than $k_0 =0.29$, imaginary part of $\omega$ vanishes and spectral density has infinitely sharp peak again.

\begin{figure} [ht!]
\begin{center}
\subfigure[]{\includegraphics[angle=0,width=0.3\textwidth]{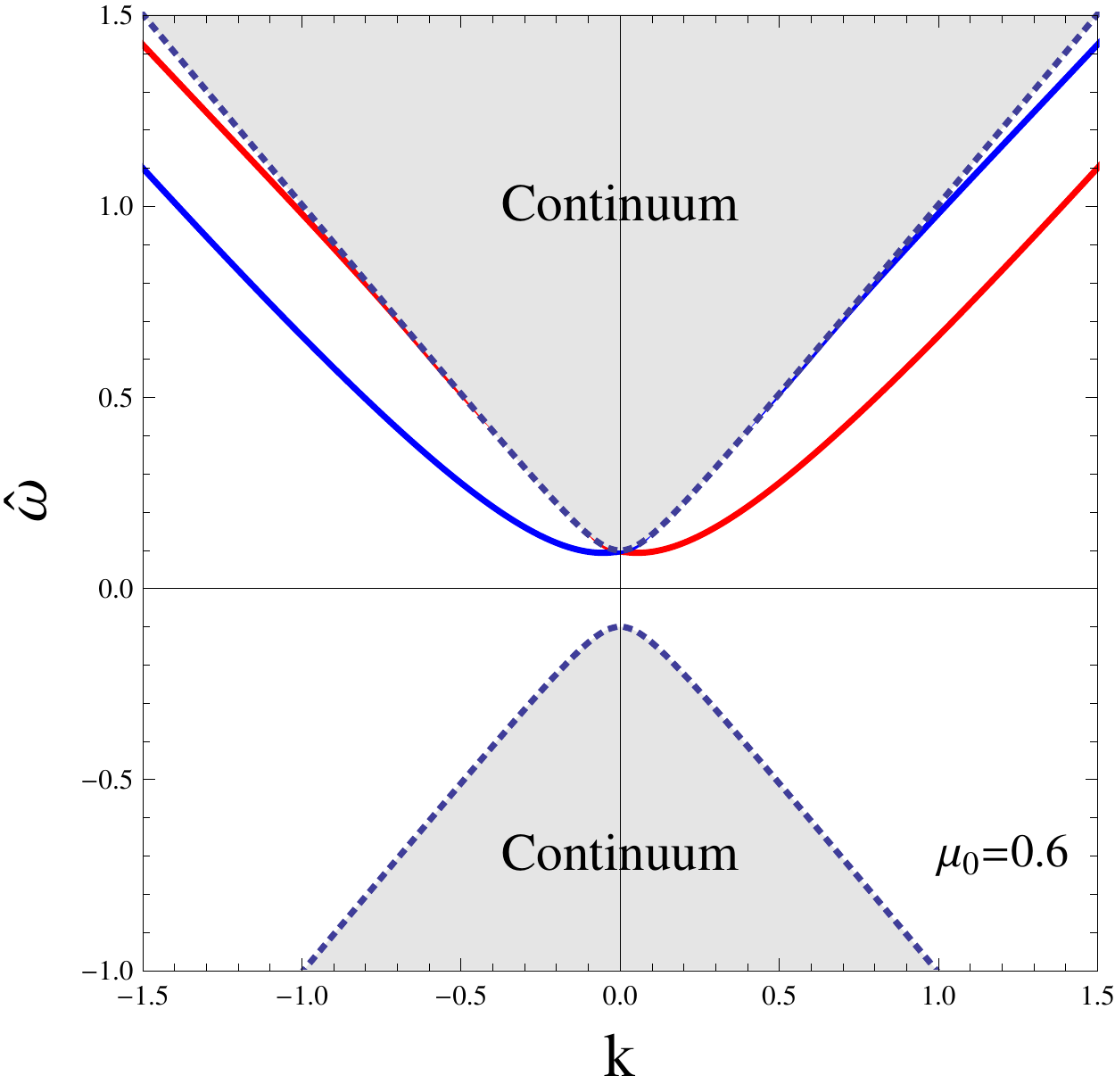}}
\hskip.5cm
\subfigure[]{\includegraphics[angle=0,width=0.3\textwidth]{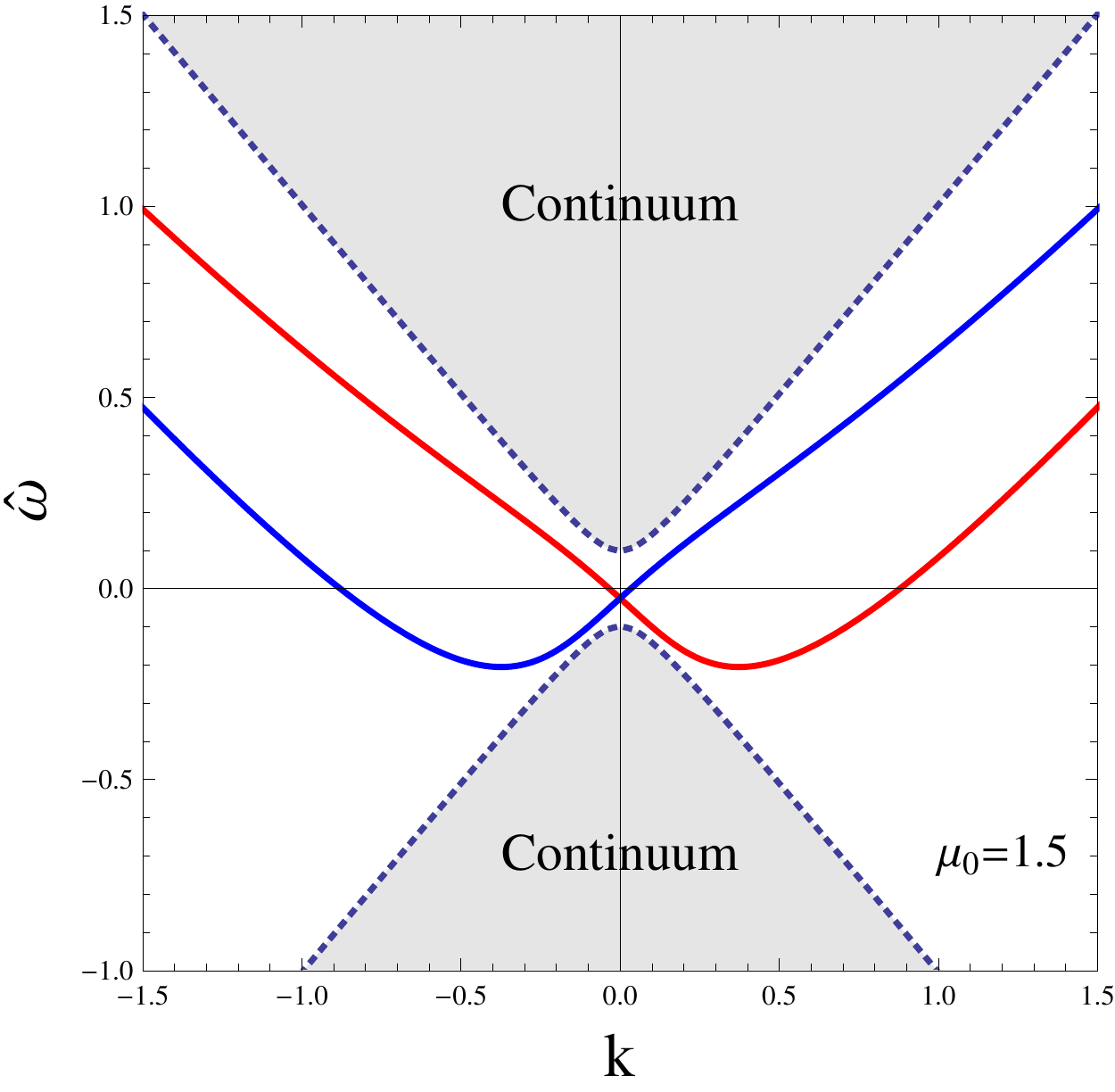}}
\hskip.5cm
\subfigure[]{\includegraphics[angle=0,width=0.3\textwidth]{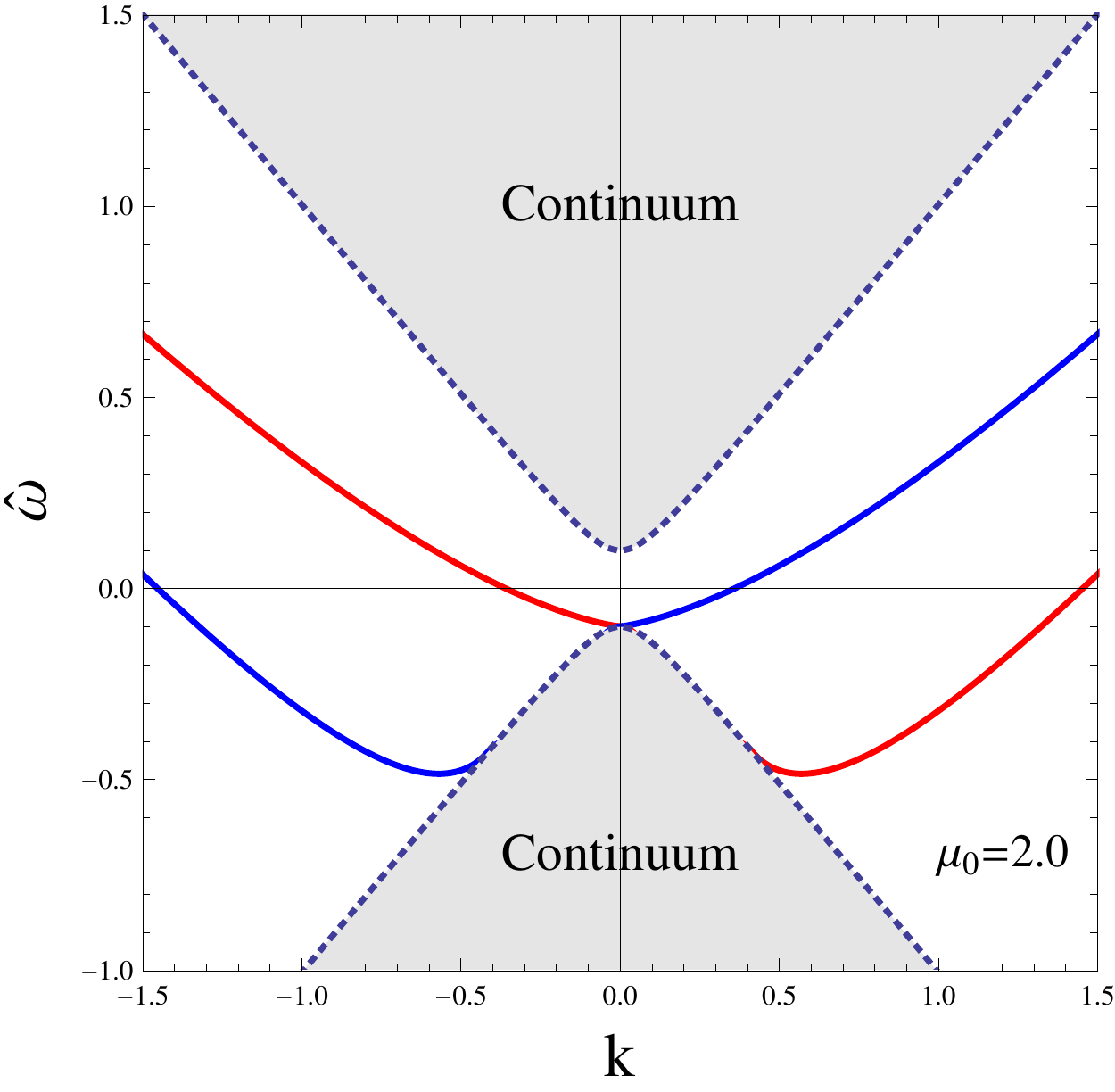}}
\caption{Density dependence of dispersion curves for $G_1^R$ and $G_2^R$. Blue solid line denotes $G_1^R$ and red line denotes $G_2^R$.\label{fig:G1G2}}
\end{center}
\end{figure}

The equation of motion (\ref{floweq2}) and IR boundary condition (\ref{bc2}) for $G^R_2$ become those of $G^R_1$ under $k \rightarrow -k$. Therefore, we can get spectral curve for $G_1^R$ by 
\be
G_1^R(k) = G_2^R (-k).
\ee 
The chemical potential dependence of the two dispersion curves are drawn in Figure \ref{fig:G1G2}.

In a recent study of the holographic thermal  fermion, Herzog et.al   \cite{Herzog:2012kx} found an interesting bulk Rashiba effect. 
Namely,   with a finite chemical potential in the boundary, one has  dual electric field in the bulk  which can couple to the fermion spin.   For  massive fermion,  one can simplify the discussion by  taking  its non-relativistic limit, which contains spin-orbit coupling. The result  is  the {\it bulk} spin-orbit coupling:
\be
H _{\pm}= \frac{k^2}{2 m_{eff}(r)} + \alpha E(r)\times\sigma  \cdot k +\dots,
\ee
where $\alpha$ is constant. 
With the different choice of spin, the splitting of dispersion relation for two fermion modes is  natural. 
One of its mode has negative slope at $k=0$, signaling the presence of plasmino in the boundary.
Therefore the dual of the bulk Rashiba effect is the nothing but  the plasmino mode generated by the density effect. 
Conversely, density induced plasmino has interesting dual interpretation as Rashiba effect. 
One should be reminded that there is no temperature induced plasmino in strong coupling regime.  
  
\subsection{Deconfined Phase}
Now we discuss the fermion dispersion relation in deconfined phase of the D4-D8 model.
The brane configuration is such that D8 and ${\bar D}$8 are decoupled and go straight to touch the horizon of the 
D4 black brane.  All the results here are given numerically and we discuss the results for massless fermion and massive fermion separately. 

\begin{figure}[htbp]
   \centering
   \includegraphics[angle=0,width=70mm]{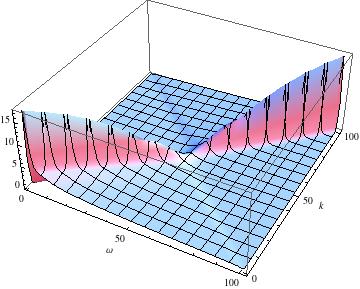} % requires the graphicx package
   \includegraphics[angle=0,width=60mm]{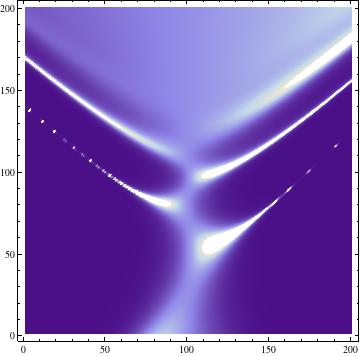} 
   \caption{  Spectral function of massless fermion.
   Left:   3D plot of spectral function at zero density figure shows vanishing thermal mass. ~Right: Density plot of spectral function at finite charge density $Q=20$. Dispersion curves of negative $k$ and positive $k$ do not meet smoothly. There is no plasmino either. Notice the $x$ axis and $y$ axis are $k$ and $\omega$ respectively. The real region for $\omega$ and $k$ is $[-5, 5]$ for both figures.}
   \label{masslessdeconfined}
\end{figure}
\begin{figure}[htbp]
   \centering
  \includegraphics[angle=0,width=60mm]{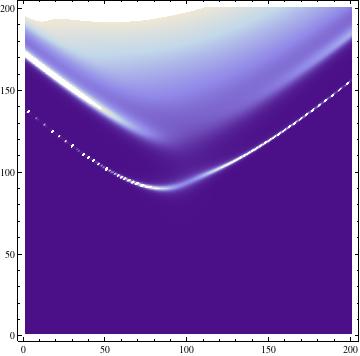} % requires the graphicx package
   \includegraphics[angle=0,width=60mm]{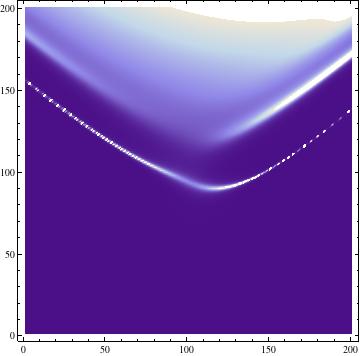} 
   \caption{Left:Density plot of spectral function of massive fermionic Green function $G^R_1$ at finite density with $m=0.5$ and $Q=30$. ~Right: Spectral function of $G^R_2$ at finite density with $m=0.5$ and $Q=30$. Plasmino exists. As density increases, dispersion curves move down along $\omega$ axis. Notice the $x$ axis and $y$ axis are $k$ and $\omega$ respectively. The real region for $\omega$ and $k$ is $[-5, 5]$ for both figures.}
   \label{fig:deconfined}
\end{figure}

Let us   discuss the massless fermion first. 
In deconfined phase, the gravity solution of D4-D8 model contains a black hole and the fermion Green function is computed by imposing the in-falling boundary conditions $G(r_H)=i$ at the horizon. Notice that when fermion is massless,   $G(r)=i$ is an exact solution for Green function at $k=0$ which also satisfies the in-falling boundary condition. This   can be noticed  from the numerical result as well. At zero density, 3D plot of spectral function (Imaginary part of retarded Green function) is shown  in the left of Figure \ref{masslessdeconfined}. There is no sharp peak and dispersion relation can be read by picking up the 
  position of the maximum heights of the spectral function. Since the dispersion curve passes through the origin, we conclude that thermal mass vanishes. 
  %Notice that at $k=0$, the spectral function  
 %$ {\rm Im} \frac{1}{\omega - v_F {k} - \Sigma(k)}  $  is not high which means that the self energy $  \Sigma(k)$ is imaginary  and its absolute value is maximum there.
 %Mastubara frequency summation factor $\tanh \frac{\omega}{2T}$  should be multiplied  to our Green function 
  %to compare it with the Real time  propagators. Notice that such Matsubara  factor, which is caused by the compactification along the imaginary time direction and is interaction %independent,  has nothing to do with the thermal mass generation which is generated by interaction.

What is the density effect? 
First, there are many dispersion relations. When both  temperature and density are present theory seems to develop 
Kaluza-Klein (KK)  tower while the temperature itself can not generate the KK tower. 
 However,  even  with  density,   plasmino is not present for zero mass case. 
 Two branches of the dispersion curve, $k<0$ and $k>0$ sector respectively, is disconnected.  

When fermion is massive,  two sectors of the dispersion curve are now connected and plasmino will be developed as we increase the density.   At zero density with finite mass, there is no plasmino. 
 As we increase  the density  the plasmino   appears at a critical density,  which is mass dependent. 
 Figure \ref{fig:deconfined} shows the plasmino at charge density $Q=30$. 
 As density increases, the dispersion curve moves down along the $\omega$ axis as it should. 

\section{Thermal mass and plasmino in Bottom Up Approach}
One may ask whether what we have seen in the previous section is model dependent or not. 
To answer this question we study the holographic fermionic spectrum from bottom up approaches.
We will see that many  features are common in both models. 

\subsection{Confined Dynamical Fermion}
The dual gravity solution describing a gapped phase is AdS soliton. 
In order to put charge source in the gravity side, we introduce dynamical fermions and take into account the coupling between 
the gauge field and fermion although we still neglect the back reaction to the gravity.  
Such fermions in hard wall model was first studied by Sachdev~\cite{Sachdev:2011ze} in  the study of holographic Luttinger theorem. 
See also~\cite{Allais:2012ye}.
We will study spectral functions and see   whether thermal mass and plasmino   exist.
   The important difference between our model and those in~\cite{Sachdev:2011ze,Allais:2012ye} is that  we use solitonic gravity background rather than hard wall. 
   For the latter, we do not have a natural boundary condition (BC) at the wall 
   while we can impose natural IR BC of fermionic fluctuations for   soliton background. 
 By ``natural" we mean that the BC comes as a necessary regularity condition at IR boundary. 
 We will see that  this boundary condition will lead us the plasmino excitations in the confined case, which is different from 
 the result of previous papers   mentioned above. 
 
We  start from the system with the action
\begin{equation}
S= \int d^{d+1} x \sqrt{-g} \left({R-2\Lambda\over 16\pi G_N} - {1\over 4e^2}F^2 + i(\bar\psi\Gamma^MD_M\psi - m\bar\psi\psi)\right)\ .
\end{equation}  
%The effective action is given by integrating out the fermion field
%\begin{equation}
%S_{eff} = \int d^{d+1} x \sqrt{-g} \left(- {1\over 4e^2}F^2 \right) + W[A]\ ,\quad e^{iW[A]} = \int D\psi \exp i(\int \bar\psi\Gamma^MD_M\psi - m\bar\psi\psi)\ .
%\end{equation} 
The five dimensional AdS soliton metric is given by 
\begin{equation}
ds^2 = r^2(-dt^2 + d\vec x^2 + f(r)dx_3) +{1\over f(r)r^2}dr^2\ ,\quad f(r):=1-{r_0^4\over r^4}\ .
\end{equation}
By factorizing $\psi = (-gg^{rr})^{-1/4}\Phi$ and choosing the following gamma matrices   basis
\[\Gamma^{\hat r} = 
\left(
\begin{array}{ccc}
 -\sigma^3 &  0 \\
 0 & -\sigma^3      
\end{array}
\right)\ ,\quad
\Gamma^{\hat t} = 
\left(
\begin{array}{ccc}
 i\sigma^1 &  0 \\
 0 & i\sigma^1      
\end{array}
\right)\ ,\quad
\Gamma^{\hat 1} = 
\left(
\begin{array}{ccc}
 -\sigma^2 &  0 \\
 0 & \sigma^2      
\end{array}
\right)\ ,\]
\[
\Gamma^{\hat 2} = 
\left(
\begin{array}{ccc}
 0 &  \sigma^2 \\
 \sigma^2 & 0      
\end{array}
\right)\ ,\quad
\Gamma^{\hat 5} = 
\left(
\begin{array}{ccc}
0 &  i\sigma^2 \\
 -i\sigma^2 & 0      
\end{array}
\right),
\]
the equation of motion for $\psi$ becomes
\begin{equation}\label{fermioneom1}
(\partial_r + {m\over r\sqrt{f}}\sigma^3)\Phi_\alpha = {1\over r^2\sqrt{f}}(i\sigma^2 (\omega+e A_t)+(-1)^\alpha k \sigma^1)\Phi_\alpha\ ,
\end{equation}where $\alpha=1,2$ are the up and down component of $\Phi$. The equation of motion for $A_t:=\phi(r)$ can be determined from Gauss's law (by variation of effective action with respect to $\phi(r)$)
\begin{equation} {1\over e^2}(\sqrt{-g}g^{tt}g^{rr}\phi'(r))' - \sqrt{-gg^{tt}}\langle \psi^\dagger\psi\rangle\ =0\ . 
%\sum_l \int {dk\over (2\pi)^2} \psi_{lk}^\dagger(r)\psi_{lk}(r)\theta(-\omega_l(k))\ ,
\end{equation}
The expectation value of $\psi^\dagger(r) \psi(r)$ can be expanded in  eigenfunctions of the Hamiltonian. For a certain background of gauge field, we require that the ground state is determined by filling all negative energy $\omega<0$ states. That is, we can rewrite the Gauss's law as
\begin{equation}\label{fluxeom}
(\sqrt{-g}g^{tt}g^{rr}\phi'(r))'- e^2\sqrt{g^{tt}\over g^{rr}}
\sum_l \int {d^2\vec k\over (2\pi)^2} \Phi_{l\vec k}^\dagger(r)\Phi_{l\vec k}(r)\theta(-\omega_l(k)) =0\ ,
\end{equation} where $l$ labels the different excited states with the same spatial momentum $\vec k$. For wave function $\Phi$, we choose the following normalization condition for each $l$ and each $\vec k$:
\begin{equation}
\int dr ~\sqrt{g^{tt}\over g^{rr}}
\Phi_{l\vec k}^\dagger(r)\Phi_{l\vec k}(r) = 1\ .
\end{equation}
Our task now is to solve (\ref{fermioneom1}) and (\ref{fluxeom}). Notice that (\ref{fluxeom}) contains the sum of the eigenstates with definite energy $\omega(k)$ and momentum $k$. $\omega(k)$ is the dispersion relation of excitations obtained from the pole of spectral function.  In order to obtain that, we first   solve the spectrum of (\ref{fermioneom1}) by an iterative method:  we  start a test flux  to compute the spectrum of (\ref{fermioneom1}), then update the flux trough (\ref{fluxeom}) and so on. We   do this until the process converges.  

\subsubsection{Boundary Conditions}
In order to solve (\ref{fermioneom1}), we need to set suitable boundary conditions for $\Phi$ and $\phi$.\\
~~~~{\it BC for Dynamical fermion}~ Following the analysis in the top down approach in the previous section, we obtain the IR boundary condition as
\be
y_2(r_0) = {-mR + \sqrt{m^2R^2+k^2-\omega^2}\over (-1)^\alpha k -\omega}\ , \quad z_2(r_0)=1,
\ee
 where $y_2$ and $z_2$ are the upper and lower component of $\Phi_2$ respectively. Notice that  the boundary value of gauge potential at the IR boundary $r=r_0$ is zero, which is consistent with the fact that there is no charge source at the IR tip of AdS soliton.
\begin{figure}[htbp]
   \centering
   \includegraphics[scale=0.8]{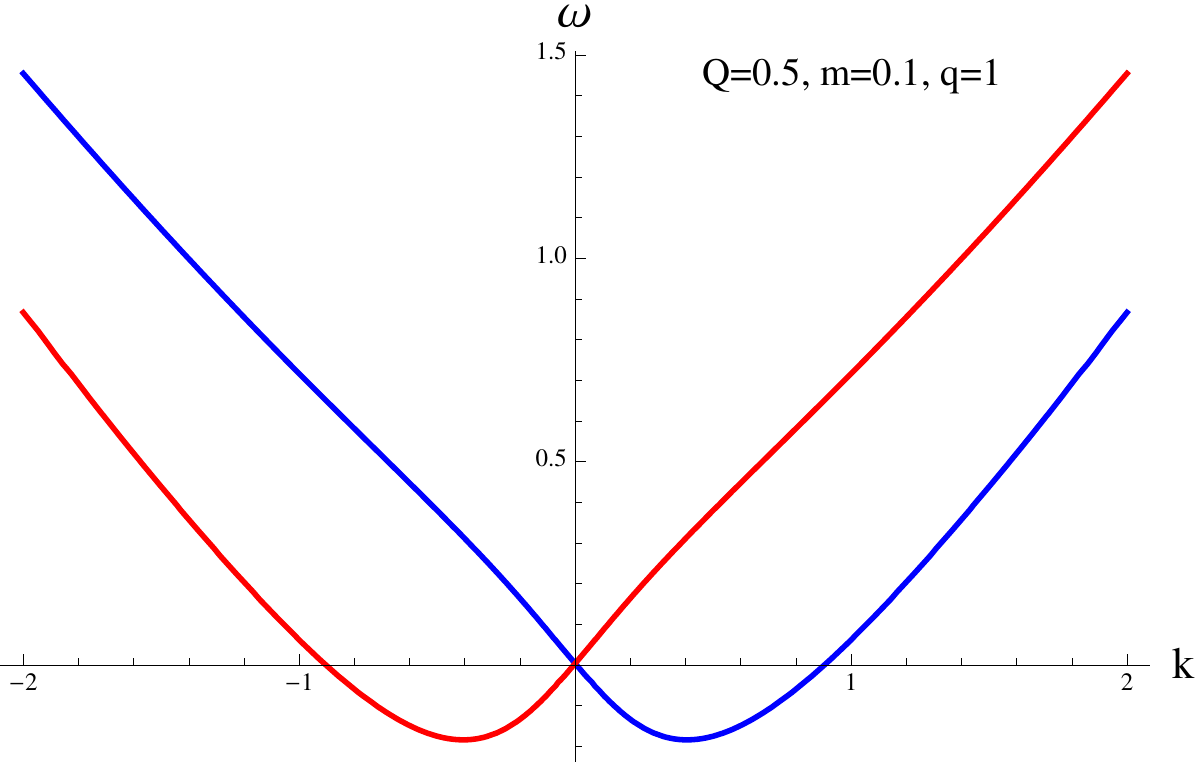} % requires the graphicx package
      \caption{Dispersion relations of two fermionic modes in the end of iteration. The upper one is normal mode coming from $G_1$ and we observe that the lower one is plasmino mode coming from $G_2$. Parameters are set to be $Q=0.5\ ,m=0.1$.}
   \label{solitondis}
\end{figure}
\begin{figure}[htbp]
   \centering
   \includegraphics[scale=0.5]{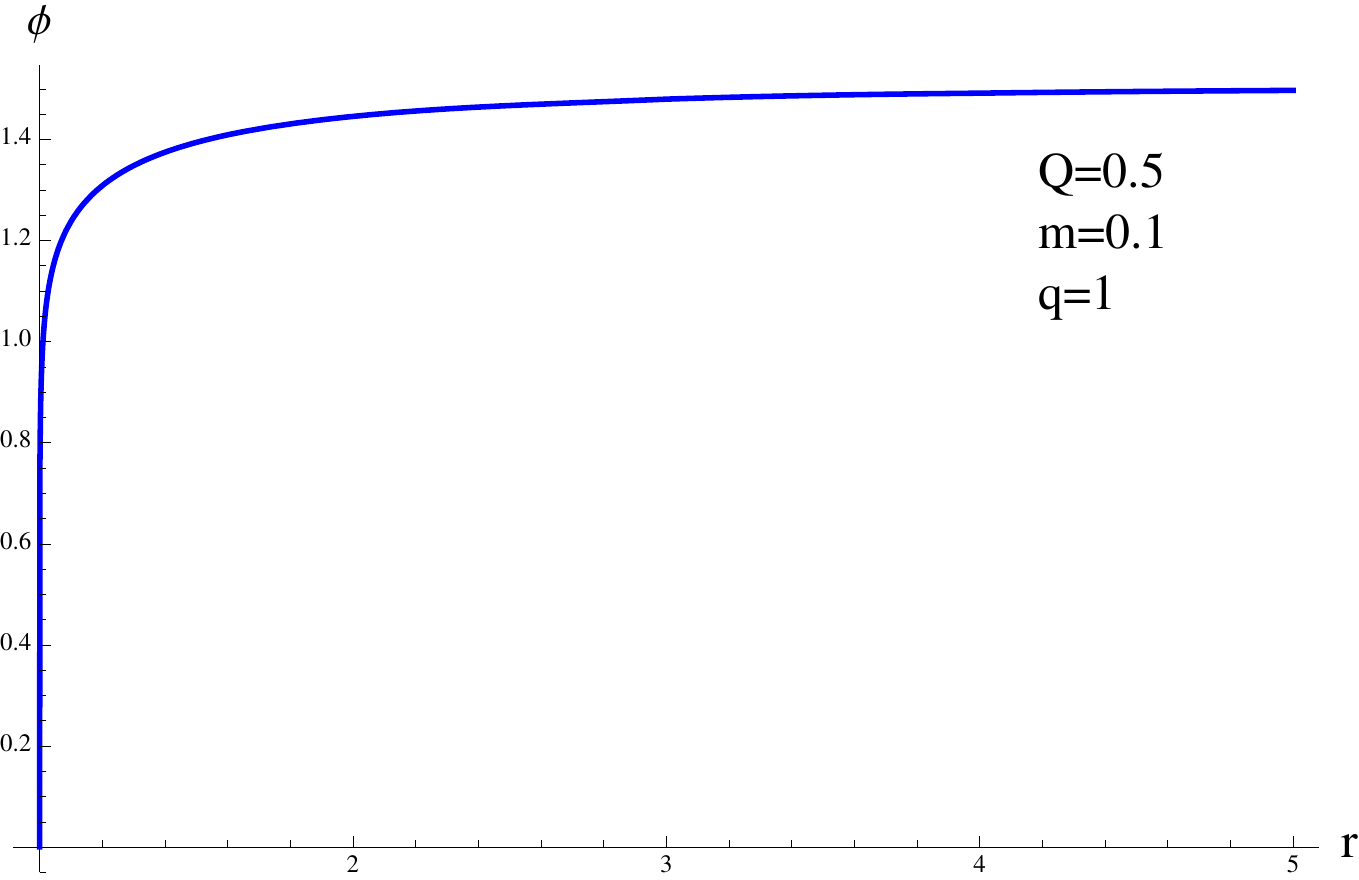} % requires the graphicx package
      \caption{%Up:~Dispersion relations of two fermionic modes in the end of iteration. The upper one is normal mode coming from $G_1$ and we observe that the lower one is plasmino mode coming from $G_2$. Parameters are set to be $Q=0.5\ ,m_5=0.1$. Down:~
      Solution of flux in the end of the iteration. The dashed line and solid line are the input flux giving the final fermion spectrum and the output of that fermion spectrum by Gauss's law. They are coincide with each other. Parameters are set to be $Q=0.5\ ,m=0.1$.}
   \label{solitonflux} 
\end{figure}

{\it BC for gauge field}~  At IR boundary the gauge field is zero as we mentioned.   The other BC we choose is the charge of the field which is a slope of the flux at the UV boundary, $Q:=r^3\phi'\Big |_{r=\infty}$. 
This is equivalent to use the canonical rather than grand canonical ensemble.  
When   we choose to fix chemical potential, we could not  achieve the convergence   during the iteration process.  
 The  dispersion relation  and the gauge field configuration are shown  in Fig \ref{solitondis} and 
Fig \ref{solitonflux} respectively for the charge density value $Q=0.5$. 
We observe that two fermionic modes exist and there is plasmino branch.
The density dependence of the dispersion relation is  similar to the top down case shown in Fig. \ref{fig:G2_mu}.

\subsection{Charged Black hole with Probe fermion}
Another simple gravity dual for a system with   density and temperature is RN-AdS black hole, where the chemical potential of a $U(1)$ current is given by boundary value of electric potential sourced by the black hole charge. A probe fermion field is expected to test the fermionic nature of such a system, for instance spectral function and Fermi  surface. This holographic model was first studied in~\cite{Lee:2008xf} and   studied further in~\cite{Liu:2009dm,Cubrovic:2009ye}. 
They  studied  spectral functions and identified fermi surface. 
In a more recent paper \cite{Herzog}, the authors studied dispersion relation of massive fermion
and study the spin structure in the bulk. It turns out that the density generated plasmino can be understood as 
spin orbit coupling in the bulk. The same phenomena continue to hold for confining case we discussed in the previous subsection.  For the zero density case, we do not have plasmino mode even for the massive fermions, not alone with 
the massless fermions. Similarly to the top down models, temperature alone does not generate any thermal mass as we will show 
below. 

To examine whether plasmino mode exists in this model,   we set up the flow equations for fermionic Green functions in RN-AdS$_5$ following the paradigm in~\cite{Liu:2009dm}, 
\be 
\sqrt{g_{ii}\over g_{rr}}\partial_r G_\alpha + 2m \sqrt{g_{ii}}G_\alpha = (-1)^\alpha k + v + ((-1)^{\alpha-1}k+v) G_\alpha^2\ ，
\ee 
where 
\be
v = \sqrt{g_{ii}\over -g_{tt}}\left(\omega + \mu q (1-{r_0^2\over r^2})\right)\ .
\ee
 At the IR  boundary, we set   $G_\alpha(r_0)=i$ due to regularity at the black hole horizon. 
 One important observation is that, the effective chemical potential is given by 
\begin{equation}
\mu_q:= \mu q\ ,
\end{equation} where $q$ is the charge of the fermion field (the coupling constant). Although $\mu$ is fixed for a given charged black hole background, we can increase $q$ in order to increase the effective chemical potential.\\

 \begin{figure}[htbp]
   \centering
   \includegraphics[angle=0,width=70mm]{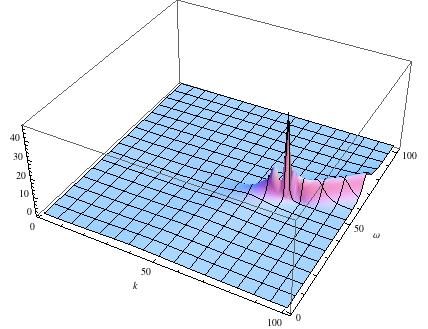} % requires the graphicx package
   \includegraphics[angle=0,width=70mm]{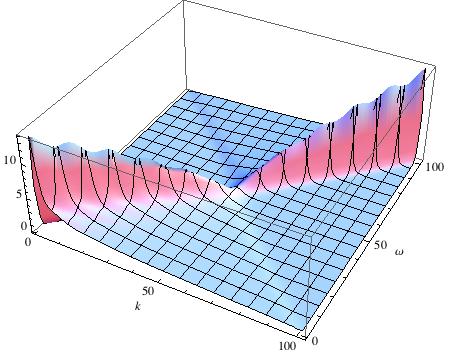} 
   \caption{ Left: For $m=0, q=1.5,Q=\sqrt{2}$, 3D plot of the spectral function. This is consistent with the spectral function obtained in~\cite{Liu:2009dm}. Right:
   For $m=0,Q=0$,  thermal mass vanishes and no plasmino exist. $Q=\sqrt{2}$ is the extremal point. Notice the real region for $\omega$ and $k$ is $[-5, 5]$ for both figures.}
   \label{dispersion0}
\end{figure}

{\it Massless fermion: }
We first show the fermionic spectral function with charge coupling $q=1$. As shown in Fig \ref{dispersion0}, when the black hole is extremal, this is precisely the result obtained in~\cite{Liu:2009dm}. For zero chemical potential, spectral function contains peaks with finite width and finite hight at $\omega=k$. This figure is qualitatively the same as the top down model. 
Therefore  our result that  no thermal mass is generated by temperature effect seems to be universal. 
 If there is no thermal mass,  there is no plasmino either.   
 
\begin{figure}[htbp]
   \centering
   \includegraphics[angle=0,width=70mm]{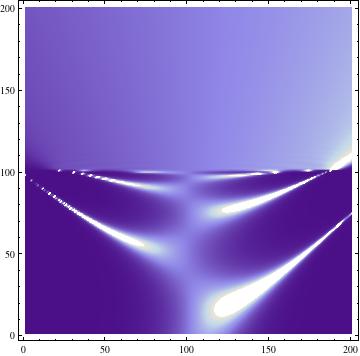} % requires the graphicx package
   \includegraphics[angle=0,width=70mm]{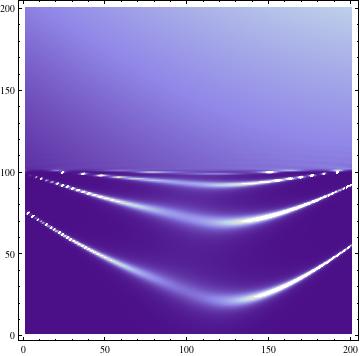} 
   \caption{ Left: $m=0, qQ=5\sqrt{2}$, there is no thermal mass and no plasmino exists for massless fermion. ~Right: $m=1, qQ=7\sqrt{2}$,   Plasmino mode exists for massive fermion if density is high enough. Notice the $x$ axis and $y$ axis are $k$ and $\omega$ respectively. The real region for $\omega$ and $k$ is $[-5, 5]$ for both figures.}
   \label{dispersion1}
\end{figure}

As we turn on and increase the density, fermionic excitations with nontrivial dispersion relations $\omega(k)$ are generated, as shown in left figure in Fig \ref{dispersion1}.
We could  not observe any  plasmino mode for zero mass for any density.

 {\it Massive fermion: }
If we turn on fermion mass, we obtain quite different results. In this case, dispersion curves are continually connected through the $k=0$ slice.
  We can observe the  plasmino mode if density is high enough.

\section{Conclusion and Discussion}
In this paper, we discussed the dispersion relation of   strongly interacting  fermions 
using the holographic method. We asked what is the field theory dual of the bulk spin-orbit coupling and 
answered that it is the density generated plasmino.
We also asked that whether the non-generation of the thermal mass for the massless fermion in finite temperature 
is universal or model dependent and answered that it is universal by working out 
them in completely different class of the theories: One in  top down and the other in the bottom up. 

We also studied the presence of the plasmino mode for various situations. 
We found that for confining geometry, 
the plasmino exists for large enough chemical potential regardless of the fermion mass.
For deconfining geometry, plasmino exists only for massive fermion with enough denisty. 
We  showed that the plasmino dispersion curve is absent at zero density, regardless of fermion mass, temperature, and 
phases (confinement/deconfinement).   It is summarized in Table \ref{table:plcondition}.

\begin{table}[ht!]
\centering
\begin{tabular}{|c|c||c|c||c|c|}
\hline
\multicolumn{2}{|c||}{Parameter}& \multicolumn{2}{c||}{Top down} & \multicolumn{2}{c|}{Bottom up} \\ \hline
                                 &                                 & Confining  & Deconfining    &  Confining   &  Deconfining   \\ \hline\hline
\multirow{2}{*}{Fermion mass}  & $=0$ &   &    &   &    \\
                                              & $> 0$ &      & $\circledcirc$ &  & $\circledcirc$ \\ \hline\hline
\multirow{2}{*}{Chemical potential}  & $<\mu_c$ &    &    &   &    \\ 
                                              & $>\mu_c$ &   $\circledcirc$      & $\circledcirc$ & $\circledcirc$ & $\circledcirc$ \\ \hline
\end{tabular}
\caption{The  conditions for existence of plasmino are indicated by the   $\circledcirc$'s. Notice that the condition is identical regardless of the  model.}                                              
\label{table:plcondition}                                              
\end{table}

From the table, we can see that the plasmino excitation is created only in the presence of chemical potential. This phenomena does not depend on the back ground geometry or method(top down or bottom up), indicating the 
 universality of  of the condition for the  presence of the plasmino modes.
Notice that temperature alone never create effective (thermal) mass, which  
however,  can be generated in the presence of large enough chemical potential. 

We also identified the dual of the bulk Rashiba effect as the presence of the boundary plasmino mode. 
Finally, it would be interesting if one   study the non-relativistic version of the squared Dirac equation and solve the radial eigenvalue equation explicitly and confirm that  the result  is identical to the spectral function calculation. 
  
 %\newpage

\section*{Acknowledgements}
This work was supported by Mid-career Researcher Program through NRF grant No. 2010-0008456. 
 It is also partly supported by the National Research Foundation of Korea(NRF) grant  through the
SRC program CQUeST with grant number 2005-0049409. The work of YS was partly supported by 
Basic Science Research Program through the National Research Foundation of Korea(NRF) funded by 
the Ministry of Education(NRF-2012R1A1A2040881).

\end{document}